\documentclass[fleqn,10pt]{olplainarticle}
 \usepackage{dirtytalk} 
\usepackage{caption}
\usepackage{csquotes}

\title{Significance of anonymity and privacy in  improving
inclusivity and diversity in Higher Education Learning Environments}

\author{Poonam Yadav\footnote{Email: poonam.yadav@york.ac.uk}}
\affil{Computer Science Department}
\affil{University of York, UK}


\begin{abstract}
Interactions between lecturers and students are the key to learning in the higher education environment. In this paper, the investigation pursues two different contexts to understand these interactions and the impact of anonymity and privacy in different interactions in the Computer Science (CS)  department. The first context \say{different interaction between a lecturer and
students} is investigated using phenomenological research approach by interviewing lecturer in CS ($N_{a}$ = 5). The second context \say{the significance of anonymity and privacy in interactions} is investigated using a quantitative and qualitative questionnaire-based research method using an online student questionnaire ($N_{b}$ = 53). The study finds a large gap between students' perception of preferred communication methods and the use of the same communication method. From the second context study, it is evident that \say{anonymity and privacy} in online surveys and module evaluations are preferred by \textit{all student participants}, thus supporting diversity and inclusivity.
\end{abstract}
\begin{document}

\maketitle

\section{Introduction}
The importance of inclusivity and diversity in Higher education has been researched significantly~\citep{Holles2021}. Diversity is a multi-dimensional concept, and its connotation depends on cultural context~\citep{Anna2012}. However, diversity is a broader aspect of student representation (including under-represented, disadvantaged or vulnerable). On the other hand, inclusivity means different to different people. \cite{Armstrong2010} provided two broad definitions that \say{inclusion is about all students with disabilities participating in all aspects of mainstream school} and \say{inclusion refers to all students actively participating in schools that value all students}. \cite{Greer2014} has provided valuable resources and information on how to create an inclusive learning environment in Higher education settings through careful course design and delivery. \cite{Greer2014} has also provided a direction to link inclusivity to students  \say{sense of belonging}. Inclusivity is a meta-construct, not something that can be empirically directly measured.
Similarly, the  \say{sense of belonging} is subjective, multi-faceted and changes over time. However, both inclusivity and   \say{sense of belonging} are directly linked with students' success and retention in higher education~\citep{Anna2012}. Effective communication and interactions among students, lecturers and staff within the higher education environment are essential for building a collaborative environment and  \say{sense of belonging}. However, communication interactions are pretty complex, especially when two key terms, Anonymity and privacy in communication, are given importance to understand the deep dynamics of interactions. The basic definition of Anonymity is \say{being without a name}. In simple terms, someone is anonymous if their identity is not known~\citep{Chauhan2015}.

The impacts of Anonymity in higher education have been researched significantly, for example, the effect of anonymous marking on students' perceptions of fairness, feedback and relationships with lecturers~\citep{Pitt2018},~\citep{Brennan2007} and how direct, or anonymous feedback and valuations from students impact the behaviour of the teacher~\citep{Corelli2015}. It is widely accepted that anonymous feedback from students provides them more authority to give honest negative feedback on the course/module evaluation~\citep{Fries2013}. However, this is only one form of interaction in HELE. 

The research question in this paper focuses on investigating one of the fundamental perspectives of interactions by taking Anonymity and privacy into consideration:  \say{What are the significance of Anonymity and privacy in improving inclusivity and diversity in Higher Education Learning Environments (HELE)?}. As described before that, Anonymity and privacy are used for a wider context in HELE; the research question in this paper is narrowed-down to the different interactions among lecturers and students to systematically analyze  \say{the significance of Anonymity and privacy associated with different interactions in improving inclusivity and diversity in HELE}. 


\section{Methodology}
The research is conducted with a systematic literature review by providing two different contexts to understand the interactions and the impact of anonymity and privacy in different interactions in the Computer Science (CS)  department. The first context, \textbf{context(A)}: \say{different interaction between a lecturer and students}, is investigated using phenomenological research approach by interviewing lecturer in CS ($N_{a}$ = 5). The second context, \textbf{context(B)}: \say{the significance of anonymity and privacy in interactions} is investigated using a quantitative and qualitative questionnaire-based research method using an online student questionnaire ($N_{b}$ = 53).
\subsection{Formulation of questions for inquiries}
The investigation pursues with understanding the detail of \say{different interaction between a lecturer and students} \textbf{context(A)} in Higher Education settings with following questions.

\begin{itemize}
    \item  What kind of interactions happens between a lecturer and a student in higher education settings?
    \item What are the purpose of interactions?
    \item How does interaction have changed in the last few years, especially during the last two years of hybrid online teaching due to Covid-19?
    \item What are the different tools used for interactions?
\end{itemize}

Further, the following questions were formulated to understand the significance of \say{anonymity and privacy in interactions} (\textbf{context (B)}).
\begin{itemize}
\item What is anonymity and privacy in higher education interactions?
\item How anonymous interactions have been established in higher education settings?
\item Why anonymous interactions are useful?
\item Are there any interactions types that benefits from being anonymous?
\item Does anonymity and privacy improve inclusivity and diversity in Higher Education Learning Environments, especially through inclusive interactions?
\end{itemize}

\subsection{Data Collection}
\begin{figure}[ht!]
\centering
\includegraphics[width=\linewidth]{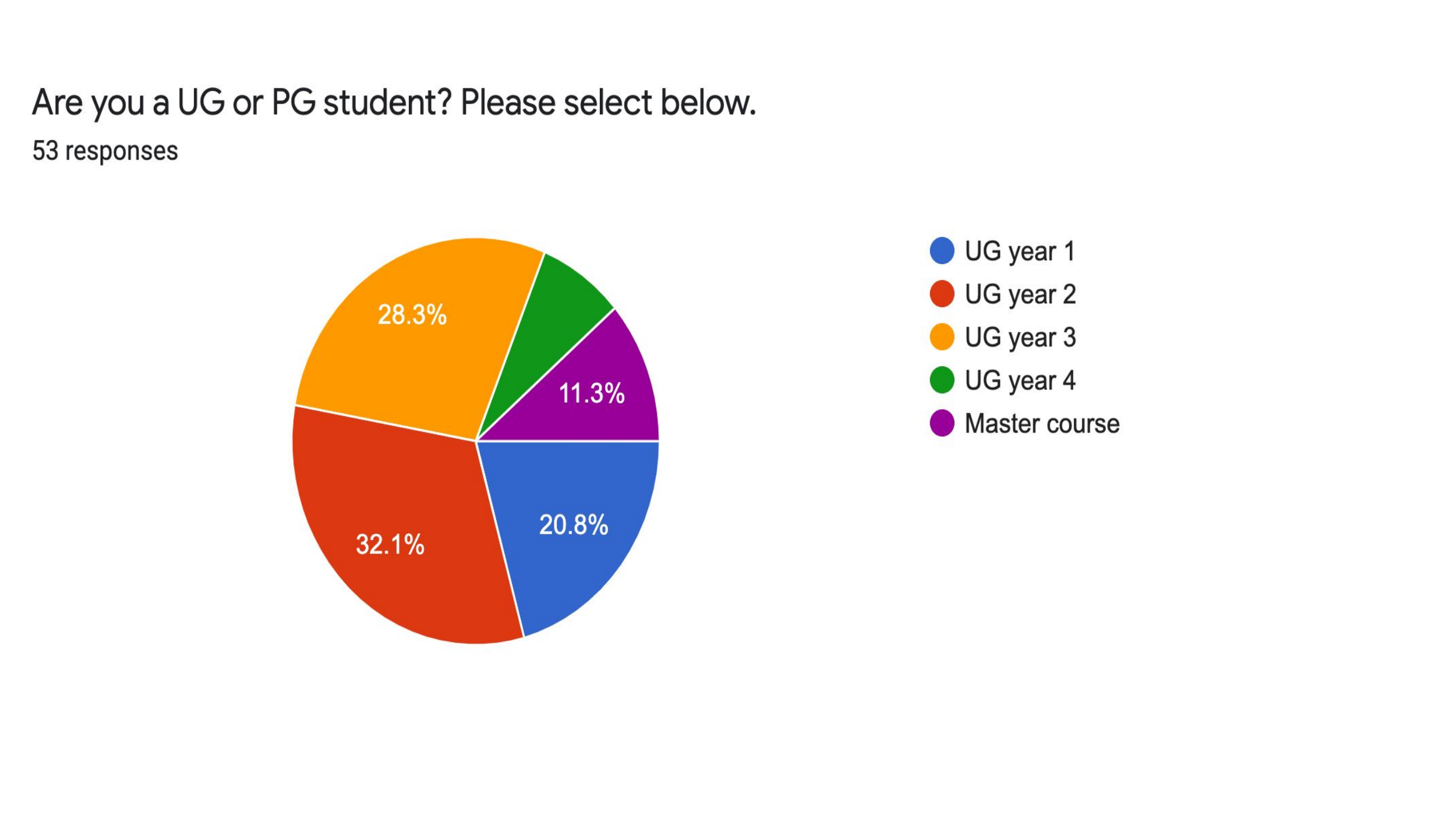}
\caption{Distribution of Survey participants based on Survey Question: Are you a UG or PG student?}
\label{fig:data1}
\end{figure}

\begin{figure}[ht!]
\centering
\includegraphics[width=\linewidth]{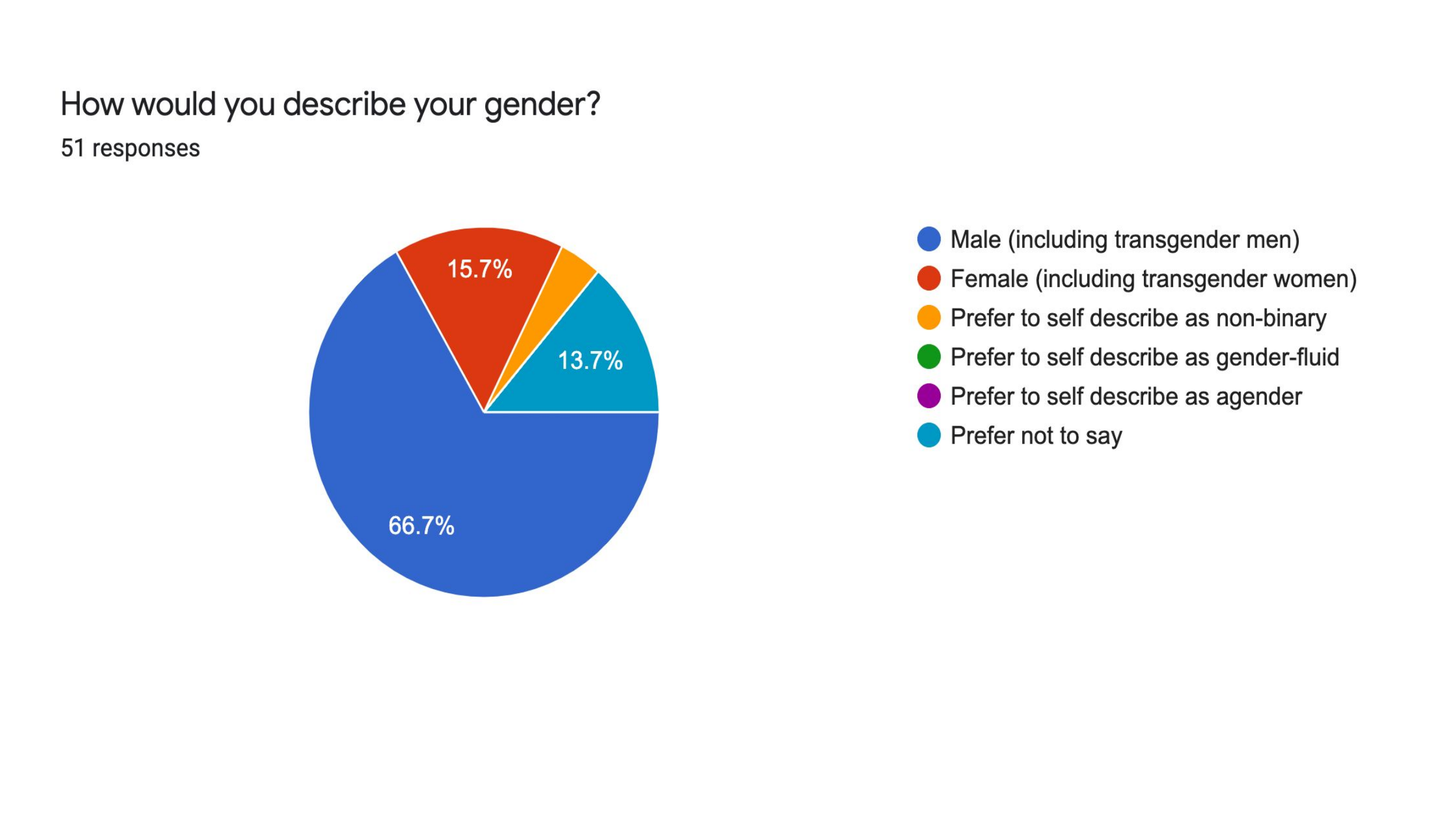}
\caption{Distribution of Survey participants based on Survey Question: How would you describe your gender?}
\label{fig:data2}
\end{figure}

The  purposeful interview-based qualitative  and survey-based quantitative and qualitative data collection methods are used for this research. 
\begin{list}{--}{}
    \item \textbf{Participants recruitment for Interviews:} The Computer Science department academic staff voluntarily participated in one-to-one interviews. The expression of interest was gathered through either direct email or the department instant messaging tool (slack).
    \item \textbf{Participants recruitment for Survey:} In the Computer Science department, all Undergraduate and Taught Postgraduate students were invited to participate in the survey. The survey (Appendix A) link was distributed through the department emailing list. A total of 52 students participated in the survey; the distribution of participants based on their study year, undergraduate level or Master's level and their gender distribution is shown in Figure~\ref{fig:data1} and Figure~\ref{fig:data2}.
\end{list}

\clearpage
\subsection{Data Storage and Integrity}
All data collected through interviews and surveys were only accessible to the interviewer and stored in a password protected University Google drive. The data collection process followed the Computer Science department Ethical Data collection, storage and use policy after taking fast-track ethical approval from the department. As a part of the policy (Appendix A), all participants were provided with all necessary information related to the project.

\subsection{Data Analysis}
The qualitative data comprises opinions and observations from the interviews and textual data gathered from the survey. The quantitative information is the statistical data collected through a google survey form. All data is used to understand two contexts - (A) Interactions between lecturers and students and (B) Anonymity and Privacy in interactions. The learning from both contexts is presented in the Conclusion section.

\section{Interactions between lecturers and students}
The research study focused on the interactions between lecturers and students in the computer science department. The data collected for this phenomenological research approach~\citep{Moustakas1994}  is through conducting one-to-one interviews with academic staff (Lecturers) and an online student survey (Appendix A). This data is analyzed to understand the different ways and settings in which Lecturers and Students interact in computer science. The following four questions were asked to all five staff who participated in one-to-one interviews. The Lecturers' responses follow each question.

\begin{list}{--}{}
\item \textbf{Q1: How do you interact with students?}\\
\textbf{Lecturer1:} \say{I frequently interact with students after lectures. Generally, 30-40 students are in the classroom, out of which 5-6 (7-8\%) students either ask questions frequently or respond to my in-classroom questions. On Modules' Virtual Learning Environment (VLE) platform, there is a discussion forum, which is non-anonymous and open to all those registered to my module who can participate by either replying to any topic or starting a new topic for discussion). I respond to issues in a discussion forum weekly.} \\
\textbf{Lecturer2:} \say{My primary interaction with students is during the lab sessions.}\\
\textbf{Lecturer3:} \say {There are nearly 70 students in my class. The lectures are pre-recorded and delivered through VLE. The main interactions with students are during synchronous Q/A sessions; only one-third of students asked questions during Q/A. I also gathered anonymous informal feedback through a google-form from students; nearly 25\% of students participated.}\\
\textbf{Lecturer4:} \say{My course is lab-based; there are nearly 30 students. The teaching material and lab experiments details are provided to students in advance. The main interactions happen during the lab sessions. Around 50\% of students ask questions during lab sessions.} \\
\textbf{Lecturer5:} \say{The course has both in-person lectures and labs. When I asked questions to students, a small fraction of them responded (4-5\%). In the labs, nearly one-third of students have some questions when prompted by me.}

\textit{\textbf{Take-away from this discussion:} From the Lecturers' responses, it is evident that primarily students (50\%) in computer science interact with the lecturer during the lab sessions. Only a tiny percentage ($<10\%$) interact in-classroom, after lecturer or through direct emails. Nearly 30-50 \% participated in online surveys or polls. When students were asked directly (through an anonymous survey) if they preferred interacting with Lecturers during or after lectures, the preference followed the normal distribution as seen in Figure~\ref{fig:data3}. Figure~\ref{fig:data4} showed the student-participants response when they were asked about their preference for interacting with lecturers through emails. The Figure shows that nearly 88\% students preferred emailing a lecturer directly. Similarly, Figure~\ref{fig:data5} shows the student-participants response when they were asked about their preference for interacting with lecturers during lab sessions. This is evident that 97\% students prefer interacting with lecturers during lab sessions.}

\begin{figure}[h]
\centering
\includegraphics[width=\linewidth]{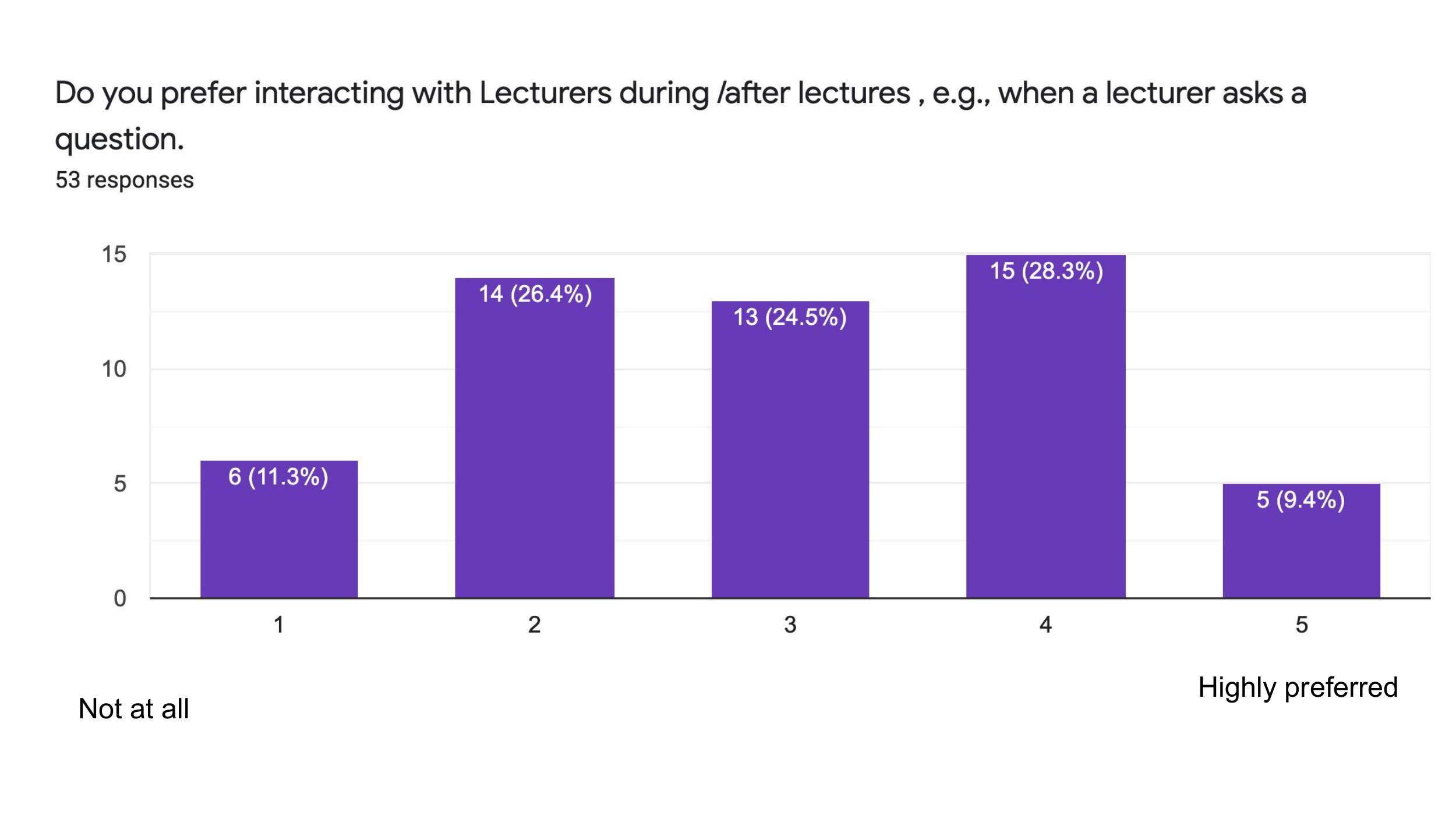}
\caption{Student-participants responded when they asked about their preference for interacting with lecturers during or after lectures (in the classroom or online). }
\label{fig:data3}
\end{figure}

\begin{figure}[ht]
\centering
\includegraphics[width=\linewidth]{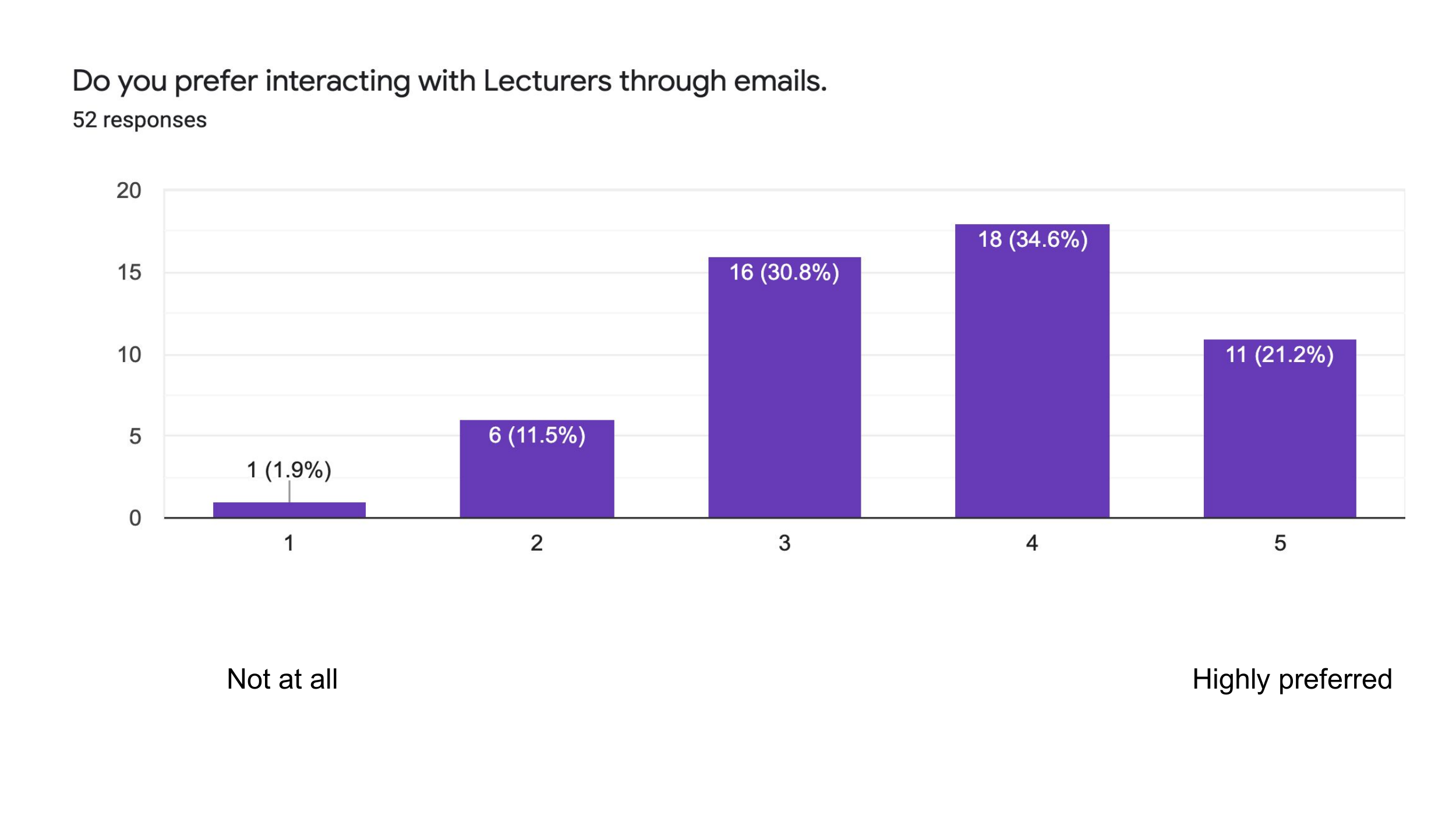}
\caption{Student-participants response when they were asked about their preference in interacting with lecturers through emails. }
\label{fig:data4}
\end{figure}

\clearpage
\begin{figure}[ht]
\centering
\includegraphics[width=\linewidth]{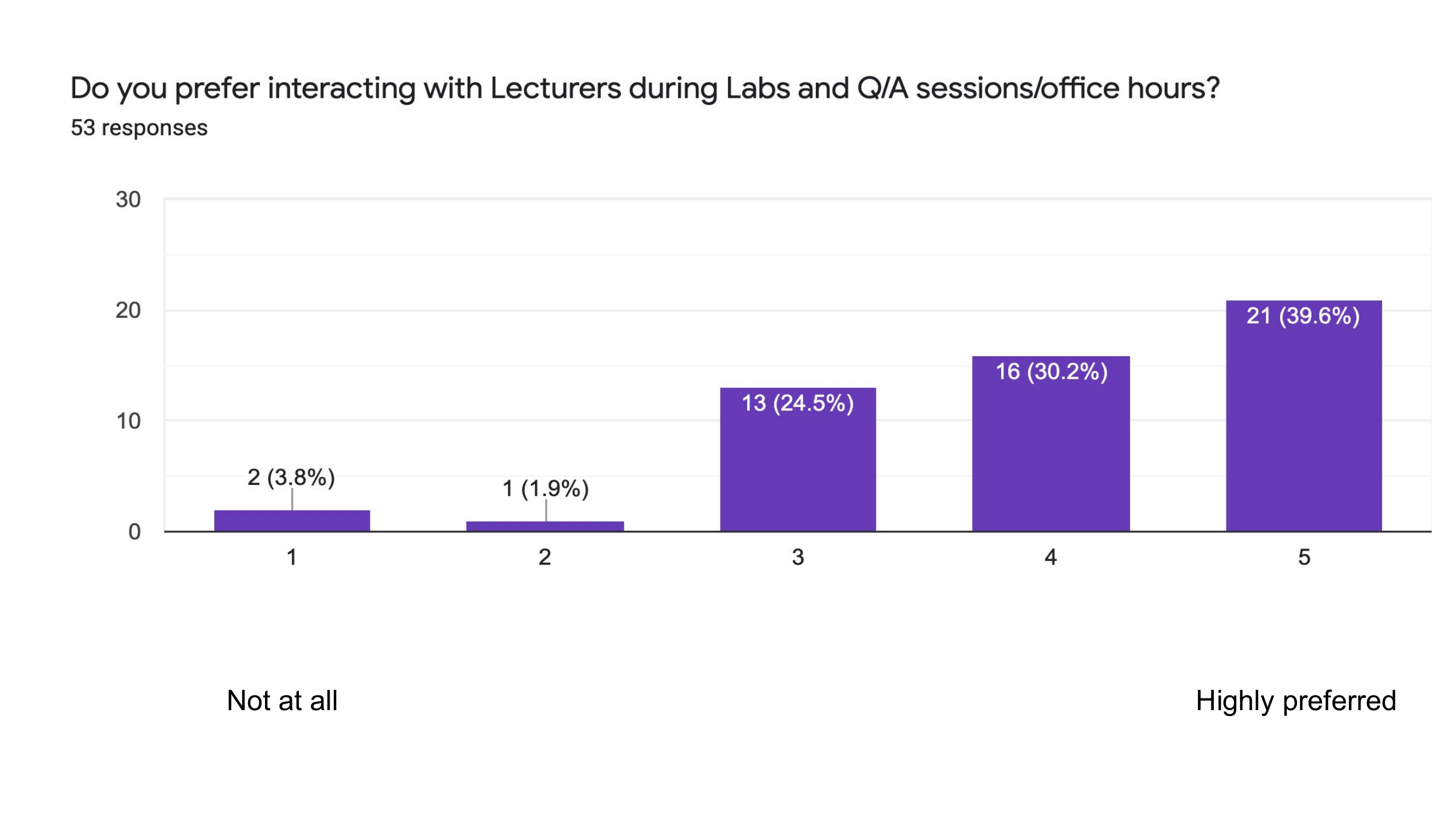}
\caption{Student-participants response when they were asked about their preference in interacting with lecturers during lab sessions.}
\label{fig:data5}
\end{figure}

\item \textbf{Q2: How do you define these interactions - real-time or non real-time?}\\
\textit{Real-time interactions are synchronous when feedback and response are shared immediately, such as the instant polls (multi-choice options) to gather instant student feedback. The non-realtime interactions are through emails, surveys/polls/feedback or evaluation forms/ questionnaires/module quizzes, where the response is not instantly provided. Students read responses/feedback on their own time ( asynchronous interaction). }

\textbf{Lecturer1:} \say{Many interactions are real-time after the lectures. Also, however, the VLE discussion forum and direct emails are asynchronous. I used Mentimeter\footnote{https://www.mentimeter.com/} to collect in-classroom synchronous anonymous responses. Nearly all students responded to polls (reasonable response rate); however, in some cases, when the poll question asked students to enter textual data in response, student participation rates dropped to the 20-50\% range. }\\
\textbf{Lecturer2:} \say{The main interactions are real-time (synchronous). However, there are occasional emails from students regarding assessments (non real-time and asynchronous).} \\
\textbf{Lecturer3:} \say{The main interactions with students are during real-time synchronous Q/A sessions.}\\
\textbf{Lecturer4: } \say{Many interactions are real-time during lab sessions}.\\
\textbf{Lecturer5:} \say{Interactions were both real-time and non real-time. The classroom questions/answers, and lab session interactions were real-time, whereas many email conversations were non real-time.}

\textit{\textbf{Take-away from this discussion:} From the Lecturers' responses, it is evident that all modules have both real-time and non-real-time interactions. Due to Covid-19, many lecturers have provided pre-recorded lectures through VLE, thus limiting real-time interactions. All lessons (in-classroom or online) could take advantage of multiple-choice polls as it emerged from the discussions that all students responded to polls ($>90\%$); however, it is an interesting observation that, in some cases, when the poll question asked students to enter textual data in response, student participation rates dropped to the 20-50\% range. }

\item \textbf{Q3: Are these interactions are lecturer initiated or student initiated?}\\
\textit{During the lecture, when the lecturer asks a direct question to the students in the classroom or the lab sessions, this interaction is \textbf{lecturer-initiated}. In contrast, the student-initiated exchange happens after lectures, labs or Q/A sessions, or drop-in sessions.}\\
\textbf{Lecturer1:} \say{Many interactions were student initiated.} \\
\textbf{Lecturer2:} \say{In labs, mostly interactions were student initiated.} \\
\textbf{Lecturer3:} \say{Many interactions are lecturer initiated.}\\
\textbf{Lecturer4:} \say{Mostly lab interactions were student initiated.}\\
\textbf{Lecturer5:} \say{In the labs, there were 70-30 \% split between  interactions which were student and lecturer initiated.}\\
\textit{\textbf{Take-away from this discussion:} From the Lecturers' responses, it is evident that many real-time interactions, in the lab or after the lecture,  are student-initiated. However, Q/A sessions and in-lecture interactions are lecturer-initiated. The asynchronous student-initiated interactions through emails are of administrating nature. However, they are related to the subject matter in the lab-based sessions.}\\
\item \textbf{Q4: Any other classroom observations, would you like to share?}\\
\textbf{Lecturer1:} \say{Students sent many administrative type emails, for example, inquiring about assessment dates or types.}\\
\textbf{Lecturer5: }\say{Due to peer motivation, students don’t like to be behind (and feel embarrassed) and thus prefer anonymous peer participation.}
\end{list}

\section{Anonymity and Privacy in interactions}
In this section, the study analyzed students' views and perceptions of anonymity and privacy in the different interactions in Computer Science from the survey data. The study further analyzed this concept concerning pedagogic research if anonymous participation helps in student engagement in improving inclusivity and diversity in HELE.\\

\begin{figure}[ht]
\centering
\includegraphics[width=\linewidth]{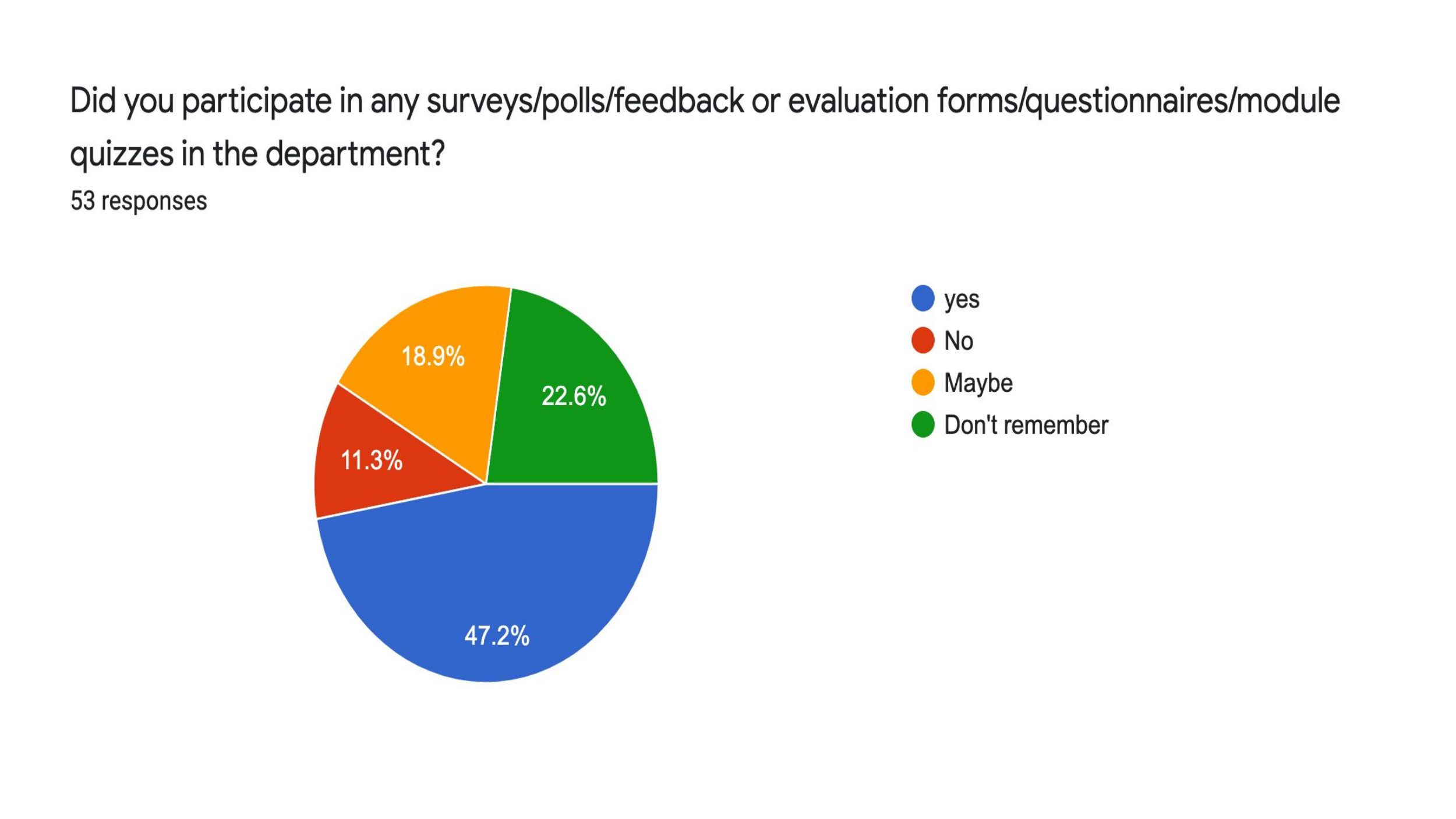}
\caption{Student-participants response when they were asked "Did you participate in any surveys/polls/feedback or evaluation forms/questionnaires/module quizzes in the department?". }
\label{fig:data6}
\end{figure}

\begin{figure}[ht]
\centering
\includegraphics[width=0.9\linewidth]{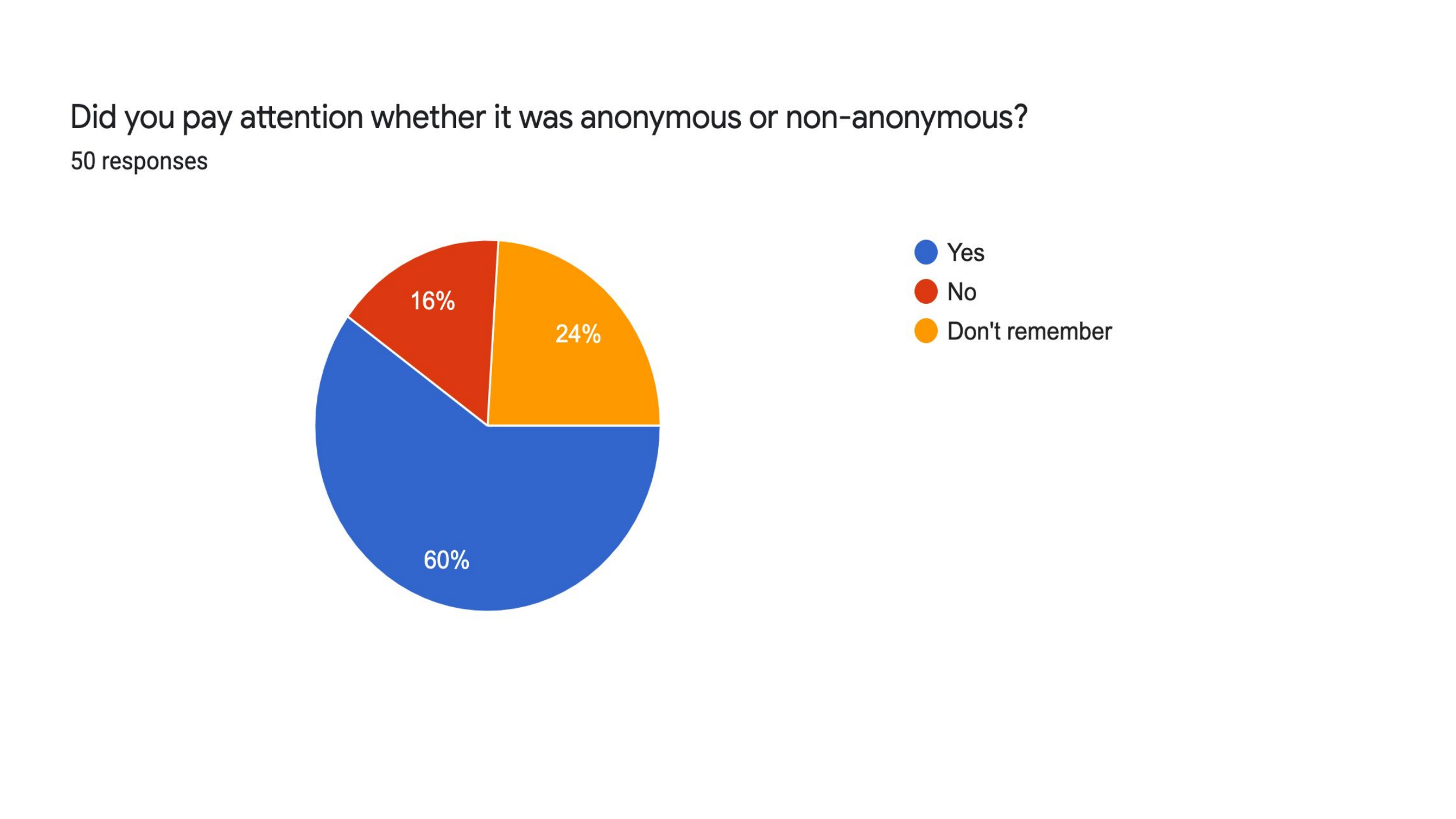}
\caption{Student participants response when they were asked "Did you pay attention whether it was anonymous or non-anonymous?". }
\label{fig:data7}
\end{figure}

\begin{figure}[ht]
\centering
\includegraphics[width=0.9\linewidth]{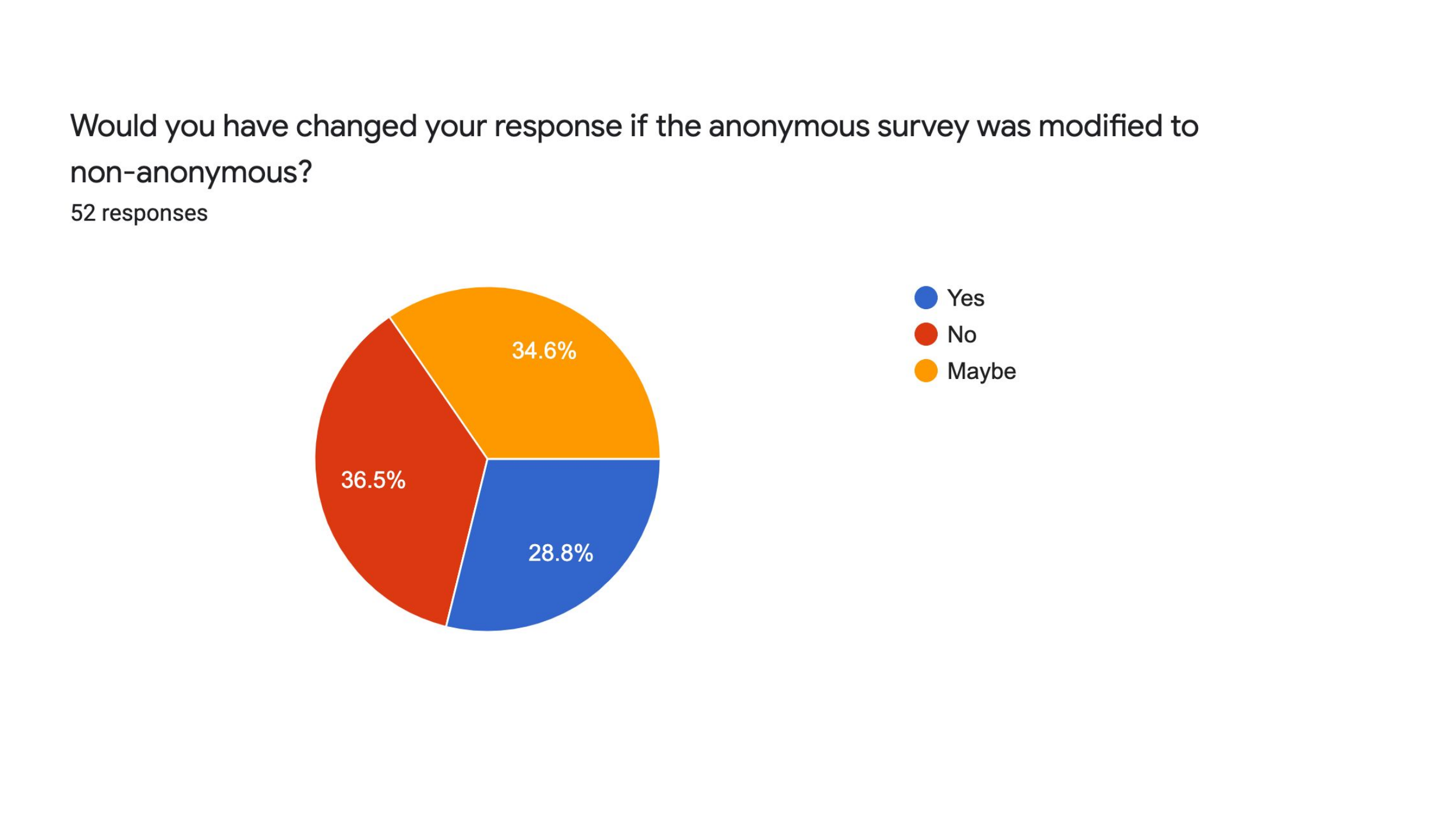}
\caption{Student-participants response when they were asked "Would you have changed your response if the anonymous survey was modified to non-anonymous?". }
\label{fig:data8}
\end{figure}

From the survey data, it is evident that nearly 48\% of students have participated in at least one interaction activity in Computer Science, as shown in Figure~\ref{fig:data6}. To understand their views on anonymity, when asked \say{Did you pay attention to whether it was anonymous or non-anonymous?}(especially in an online context). 60\% of the students paid attention to the interaction method, whether it was anonymous or non-anonymous, as shown in Figure~\ref{fig:data7}. Further on the same topic when asked if they would change their response \say{if  the anonymous survey was modified to non-anonymous?}, as a response to this question, 28.8\% students responded \say{yes}, 36.5\% students responded \say{no} and 34.6\% students responded \say{maybe} as shown in Figure~\ref{fig:data8}. When looking at other questions' responses in the questionnaire (Appendix A), this is evident that students only considered module evaluation and feedback forms as a mode of communication interaction for answering the above questions and based their above responses on based of this. The impact of privacy and anonymity in other interactions (direct emails or face-to-face lab and lecture interactions) were not analyzed due to insufficient data gathered through the questionnaire.

\clearpage
\textit{\subsection*{Concerns around non-anonymous participation}}

\begin{figure}[ht]
\centering
\includegraphics[width=\linewidth]{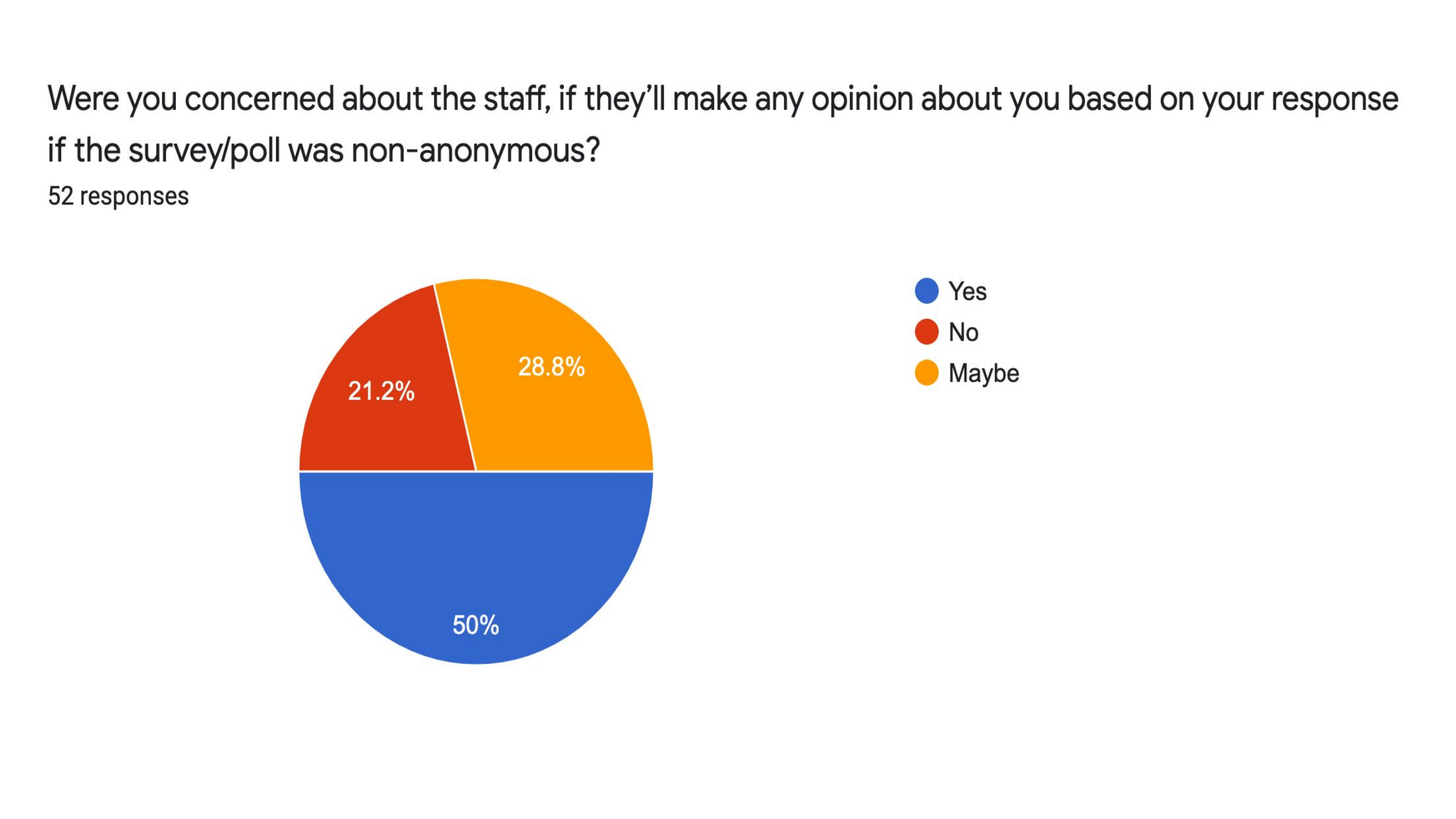}
\caption{Student participants response when they were asked "Were you concerned about the staff, if they’ll make any opinion about you based on your response if the survey/poll was non-anonymous?". }
\label{fig:data9}
\end{figure}

\cite{Kanagasabai2013} surveyed the impact of anonymity on student posting behaviour on an online forum and found students are more likely to post to discussion boards when anonymous posting is enabled than when identified posting is required. \say{A combination of individual-level factors, including online privacy concern, self-consciousness and self-efficacy, were predictive of the likelihood of making identified postings, but only self-efficacy was a significant unique predictor of anonymous postings}\citep{Kanagasabai2013}. To conform with this findings, two questions were asked to students to get their concerns regarding non-anonymous participation \say{Were you concerned about the staff, if they’ll make any opinion about you based on your response if the survey/poll was non-anonymous?}. As a response, only 21.2 \% responded \say{no}, means they were comfortable being identified without any concern of judgment by the lecturer or staff, whereas 78.8 \% responded the \say{yes or maybe}, highlighting their fears as shown in Figure~\ref{fig:data9}.

\begin{figure}[ht]
\centering
\includegraphics[width=\linewidth]{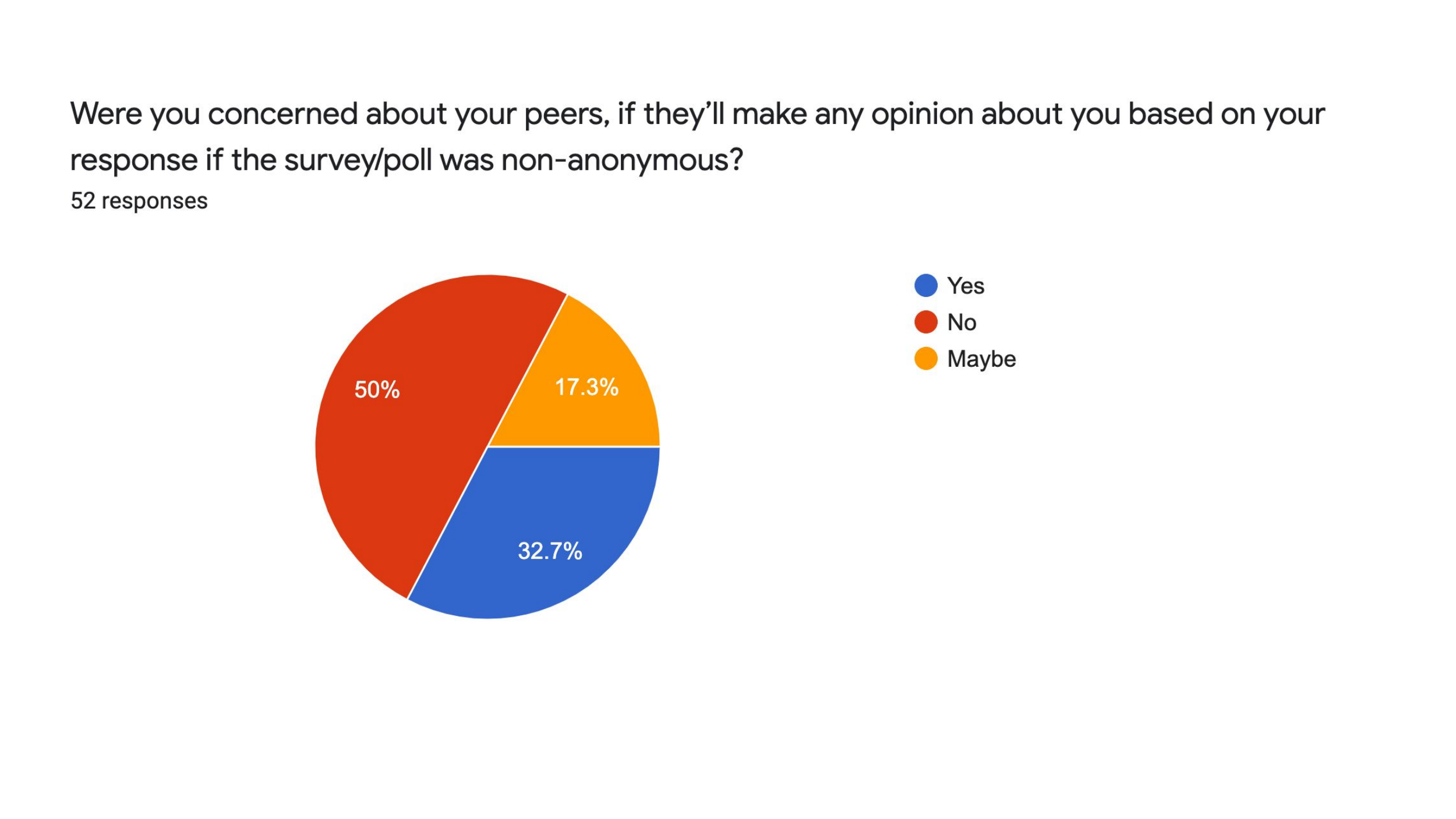}
\caption{Student participants response when they were asked "Were you concerned about your peers, if they’ll make any opinion about you based on your response if the survey/poll was non-anonymous?". }
\label{fig:data10}
\end{figure}

Another question was asked \say{Were you concerned about your peers, if they'll make any opinion about you based on your response if the survey/poll was non-anonymous?}, as a response, 50\% responded \say{no}, means they were comfortable being identified without any concern of judgment by their peers, whereas 50 \% responded the \say{yes or maybe}, highlighting their fears as shown in Figure~\ref{fig:data10}. 

Interestingly, all-female students (including transgender women) were concerned about the lecturers and peers making judgments based on their non-anonymous responses. Therefore, it is evident that feedback/evaluation based surveys should be anonymous for inclusivity and diversity. The other exciting comment provided by a student participant -\say {In general, I'd expect anonymous surveys to be better given that some students might have their honest opinions affected or remain unvoiced if they have to provide non-anonymous answers. That said, I think back-and-forth communication is essential, so even anonymous surveys should have an optional section where students could provide an email / self-identify to be reached out to in case the surveyors would like to ask further questions.}. This has opened up an exciting direction to investigate further how to incorporate either an option to self-identify mechanism or a back-and-forth anonymous discussion platform for anonymous surveys, which will benefit both students and lecturers~\citep{Tan2020}. 

\section{Conclusions}
In the computer science department, it is evident that most students (50\%) interact with the lecturer during the lab sessions. However, 97\% showed this is their preferred way of interaction. It means that even though all students prefer lab interactions, only half of them use this method for interaction. Similarly, nearly 88\% students chose to email a lecturer directly. However, there is only a tiny percentage (< 10\%) who interact in-classroom, after lectures or through direct emails. According to lecturers, nearly 30-50 \% participated in online surveys or polls, which is supported by student survey data that  48\% of students participated in at least one interaction activity (surveys, polls, feedback, evaluation forms, questionnaires, module quizzes). In terms of anonymous survey and feedback, 50\% were concerned about the lecturer making an opinion about them compared to 32.7\% who were worried about their peers. However, 97\% preferred in-lab direct interactions without worrying about the lecturers and peers making any opinions or views.
\bibliography{sample}
\newpage
\clearpage
\section*{Appendix A}

\begin{figure}[ht]
\centering
\includegraphics[page = 1, width=10cm]{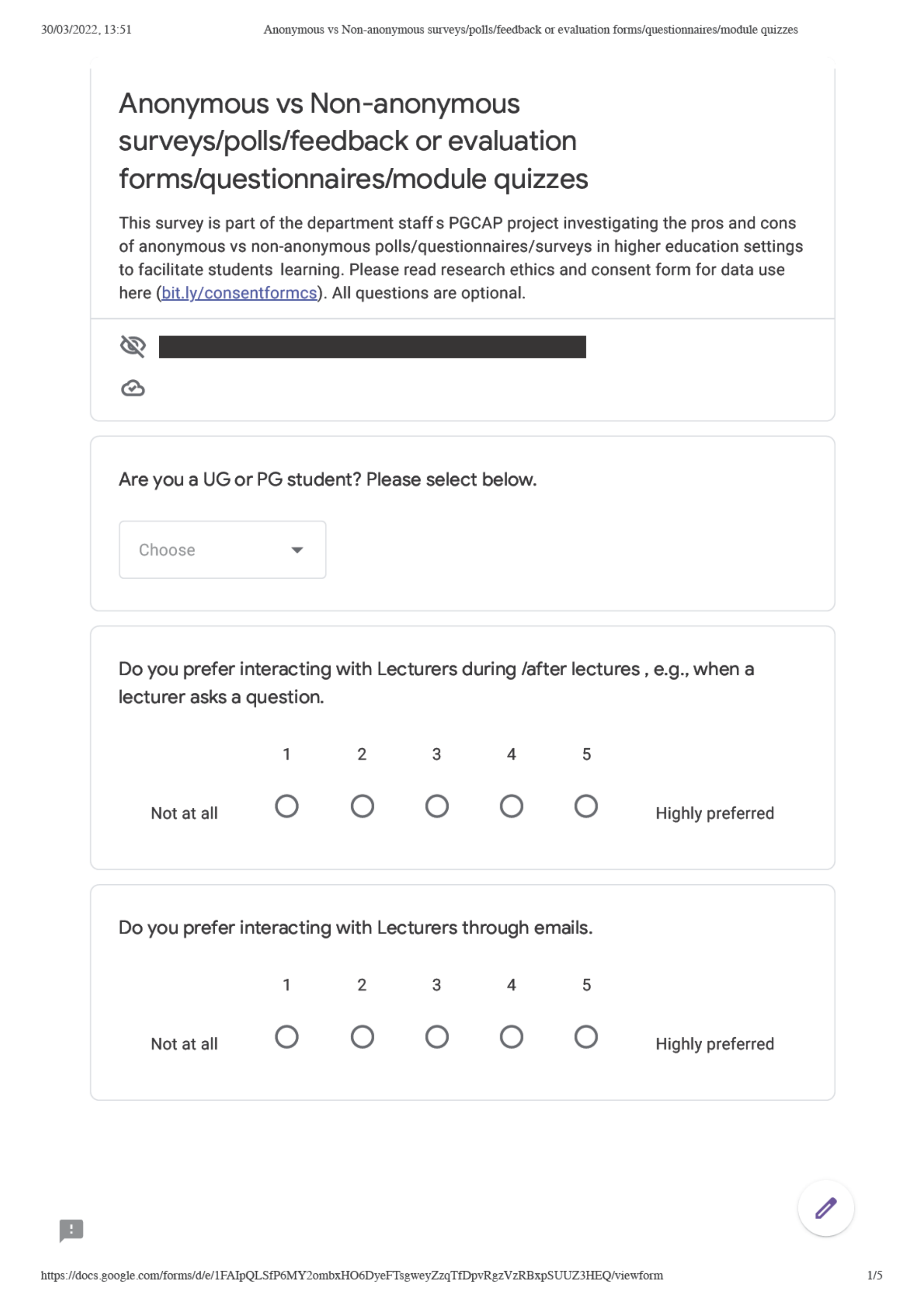}
\caption*{Page 1 of the survey used for collecting responses from student participants. }
\end{figure}

\begin{figure}[ht]
\centering
\includegraphics[page = 2, width=10cm]{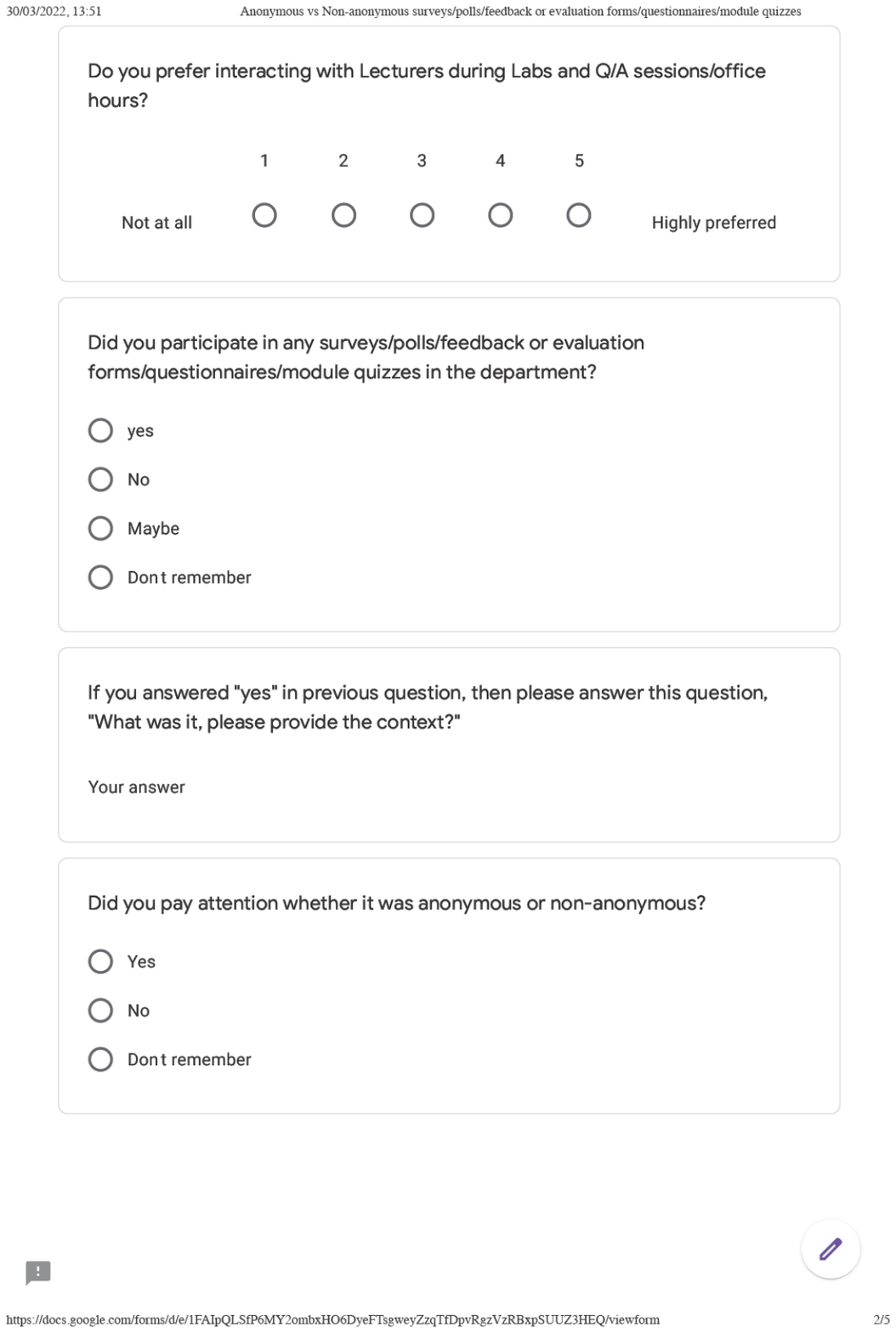}
\caption*{Page 2 of the survey used for collecting responses from student participants. }
\end{figure}

\begin{figure}[ht]
\centering
\includegraphics[page = 3, width=10cm]{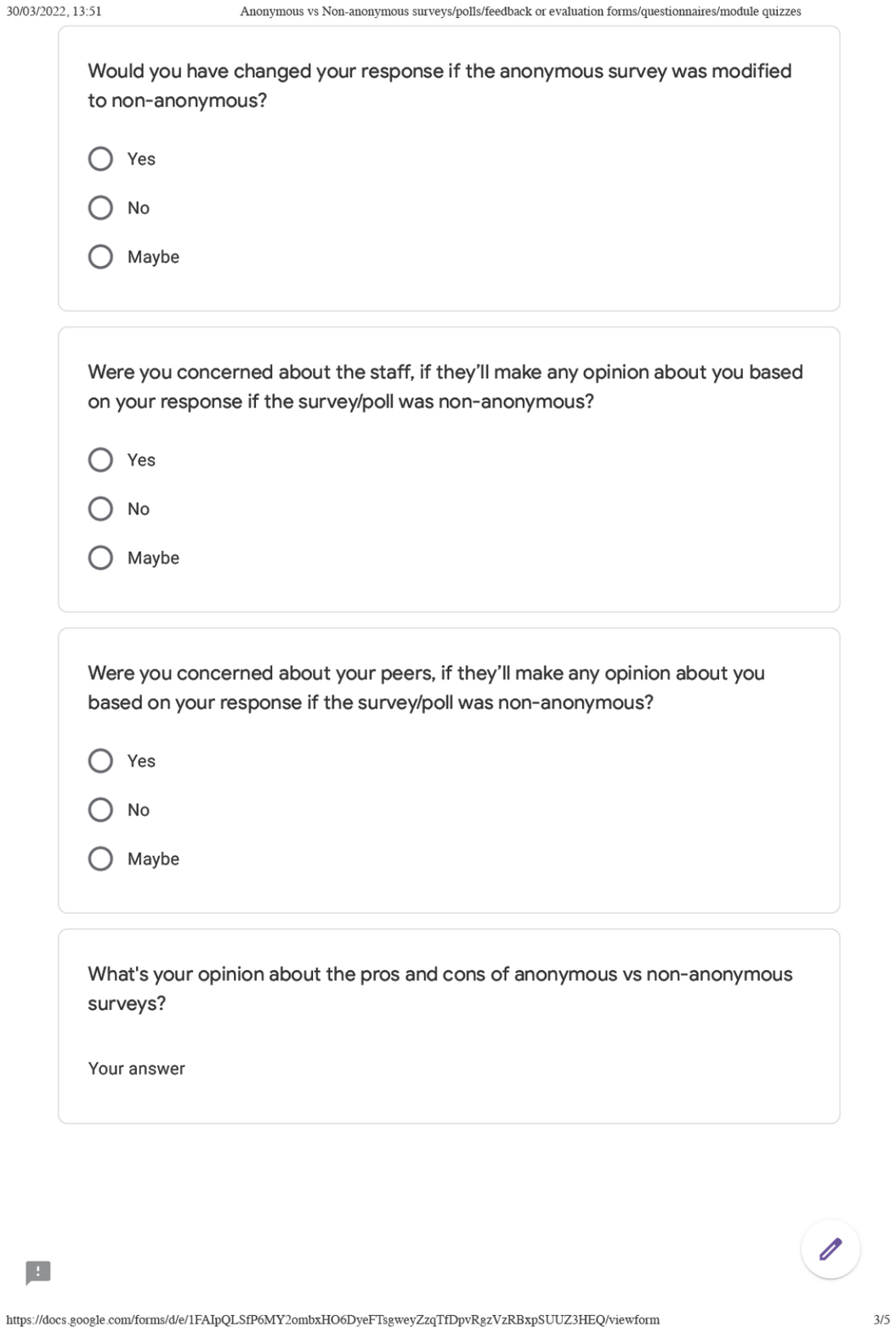}
\caption*{Page 3 of the survey used for collecting responses from student participants. }
\end{figure}

\begin{figure}[ht]
\centering
\includegraphics[page = 4, width=10cm]{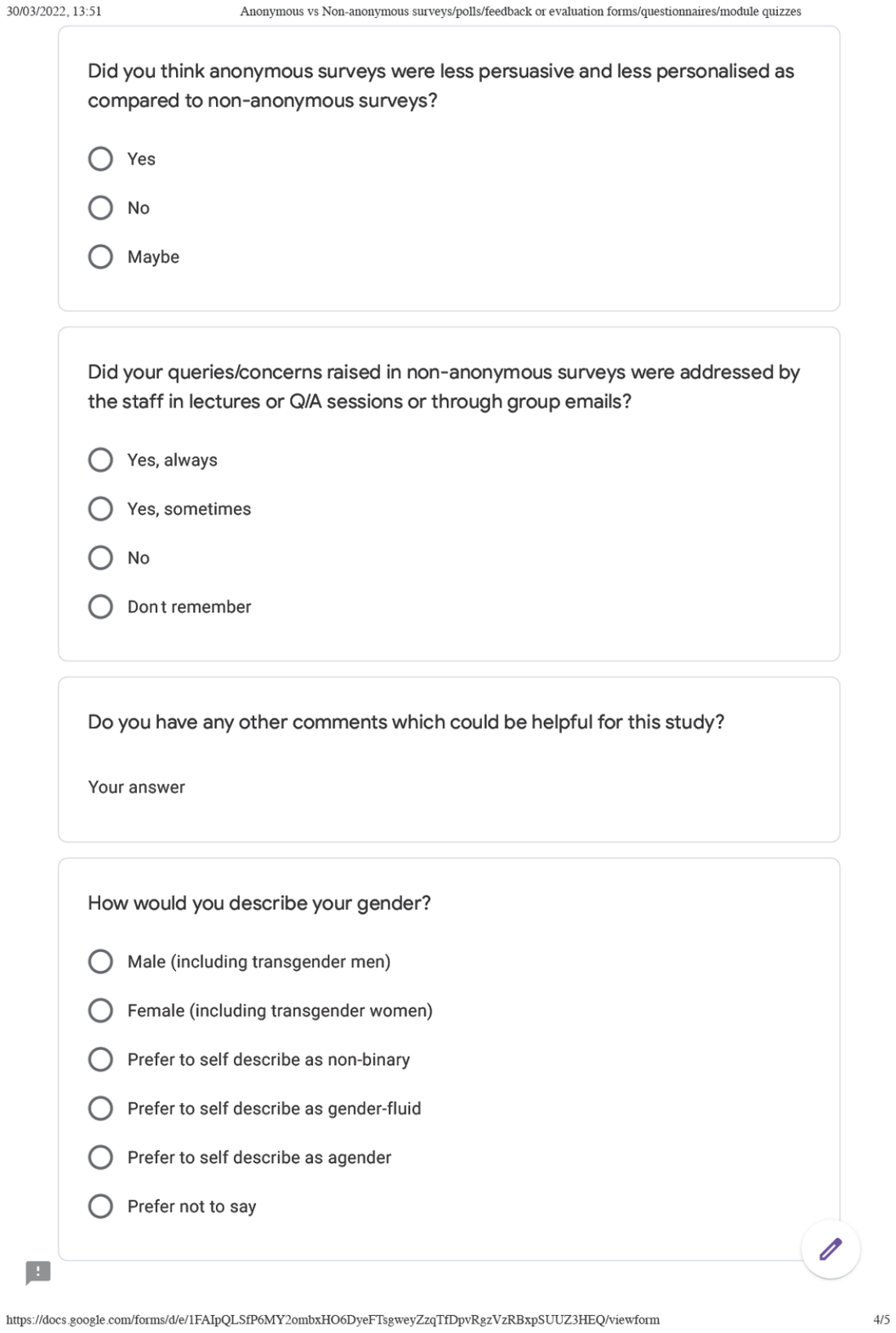}
\caption*{Page 4 of the survey used for collecting responses from student participants. }
\end{figure}

\end{document}